\documentclass[twocolumn,showpacs,preprintnumbers]{revtex4}
\usepackage{graphicx}
\usepackage{dcolumn}
\usepackage{bm}
\begin{document}
\newcommand{\ds}{\displaystyle}
\newcommand{\be}{\begin{equation}}
\newcommand{\en}{\end{equation}}
\newcommand{\bea}{\begin{eqnarray}}
\newcommand{\ena}{\end{eqnarray}}
\topmargin -2cm
\title{Self--dual Lorentzian wormholes in n--dimensional Einstein gravity}
\author{Mauricio Cataldo}
\altaffiliation{mcataldo@ubiobio.cl} \affiliation{Departamento de
F\'\i sica, Facultad de Ciencias, Universidad del B\'\i o--B\'\i
o, Avenida Collao 1202, Casilla 5-C, Concepci\'on, Chile.}
\author{Patricio Salgado}
\altaffiliation{pasalgad@udec.cl} \affiliation{Departamento de
F\'\i sica, \\ Facultad de Ciencias F\'\i sicas y Matem\'aticas,
Universidad de Concepci\'on, Casilla 160-C, Concepci\'on, Chile.}
\author{Paul Minning}
\altaffiliation{pminning@udec.cl} \affiliation{Departamento de
F\'\i sica, Facultad de Ciencias F\'\i sicas y Matem\'aticas,
Universidad de Concepci\'on, Casilla 160-C, Concepci\'on, Chile.
\\}
\date{\today}
\begin{abstract}
{\bf {Abstract:}}  A family of spherically symmetric, static and
self--dual Lorentzian wormholes is obtained in n--dimensional
Einstein gravity. This class of solutions includes the
n--dimensional versions of the Schwarzschild black hole and the
spatial--Schwarzschild traversable wormhole. Using isotropic
coordinates we study the geometrical structure of the solution,
and delineate the domains of the free parameters for which
wormhole, naked singular geometries and the Schwarzschild black
hole are obtained. It is shown that, in the lower dimensional
Einstein gravity without cosmological constant, we can not have
self--dual Lorentzian wormholes.

\vspace{0.5cm} 
\end{abstract}
\smallskip\
\maketitle \preprint{APS/123-QED}
\section{Introduction}
During the past two decades, considerable interest has grown in
the field of wormhole physics, which has rapidly grown into an
active area of research. Two separate directions emerged: one
relating to Euclidean signature metrics and the other concerned
with Lorentzian ones.

From the point of view of the Euclidean path integral formulation
of quantum gravity Coleman\cite{Coleman}, and Giddings and
Str\"{o}minger~\cite{Strominger}, among others, have shown that
the effect of wormholes is to modify the low energy coupling
constant, and to provide probability distributions for them.

On the purely gravitational side, the interest has been focused on
Lorentzian wormholes. The interest in this area was especially
stimulated by the pioneering works of Morris, Thorne and
Yurtsever~\cite{Morris,Morris1}, where static, spherically
symmetric Lorentzian wormholes were defined and considered to be
an exciting possibility for constructing time machine models with
these exotic objects, for backward time travel (see
also~\cite{Novikov}).

Most of the efforts are directed to study static configurations
that must have a number of specific properties in order to be
traversable. The most striking of these properties is the
violation of energy conditions. This implies that the matter that
generates the wormholes is exotic~\cite{Morris,Morris1,Visser},
which means that its energy density is negative, as seen by static
observers. These wormholes have no horizons and thus allow
two--way passage through them.

On the other hand, many theories of unification (e.g. superstring
theories, M--theory) require extra spatial dimensions to be
consistent. Until today a number of important solutions of
Einstein equations in higher dimensions have been obtained and
studied, and they have led to important generalizations and wider
understanding of gravitational fields. For example, Myers and
Perry~\cite{Myers} have found the n--dimensional versions of the
Schwarzschild, Reissner--Nordstrom and Kerr solutions, and have
discussed the associated singularities, horizons and topological
properties. Dianyan~\cite{Dianyan} has considered the
n--dimensional Schwarzschild--de Sitter and
Reissner--Nordstrom--de Sitter spacetimes.

An exact general solution of Einstein equations for spherically
symmetric perfect fluids in higher dimensions was found by Krori,
Borgohain and Das~\cite{Krori}. Ponce de Leon and
Cruz~\cite{Norman} discuss the question of how the number of
dimensions of space--time influences the equilibrium
configurations of stars.

Higher dimensional wormholes have also been considered by several
authors. Euclidean wormholes have been studied by Gonzales--Diaz
and by Jianjun and Sicong~\cite{Gonzales} for example. The
Lorentzian ones have been studied in the context of the
n--dimensional Einstein--Gauss--Bonnet theory of
gravitation~\cite{Biplab}. Evolving higher dimensional wormholes
have been studied by Kar and Sahdev and by A. DeBenedictis and
Das~\cite{Sayan,DeBenedictis}.

It is the purpose of the present paper to obtain, following the
prescription provided by Dadhich, Kar, Mukherjee and
Visser~\cite{Dadhich}, a family of spherically symmetric, static
and self--dual Lorentzian wormholes in n--dimensional Einstein
Gravity.

The outline of the present paper is as follows: In Sec. II we
briefly review some important aspects of 4--dimensional Lorentzian
wormholes and give the definition of self--dual wormholes
developed in~\cite{Dadhich}. In Sec. III a new class of metrics is
presented which represent self--dual Lorentzian wormholes in
n--dimensional Einstein gravity. Their properties are studied. We
use the metric signature ($-+++$) and set $c=1$.

\section{4--dimensional Lorentzian wormholes}
\subsection{Characterization of Lorentzian wormhole}
The metric ansatz of Morris and Thorne~\cite{Morris} for the
spacetime which describes a static Lorentzian wormhole is given by
\begin{eqnarray}\label{4wormhole}
ds^2=-e^{2\phi(r)}dt^2+\frac{dr^2}{1-\frac{b(r)}{r}}+r^2(d\theta^2+sin^2
\theta d \varphi^2),
\end{eqnarray}
where the functions $\phi(r)$ and $b(r)$ are referred to as
redshift function and shape function respectively.

Morris and Thorne have discussed in detail the general constraints
on the functions $b(r)$ and $\phi(r)$ which make a
wormhole~\cite{Morris}:

Constraint 1: A no--horizon condition, i.e. $e^{\phi(r)}$ is
finite throughout the space--time in order to ensure the absence
of horizons and singularities.

Constraint 2: Minimum value of the r--coordinate, i.e. at the
throat of the wormhole $r=b(r)=b_0$, $b_0$ being the minimum value
of $r$.

Constraint 3: Finiteness of the proper radial distance, i.e.
\begin{eqnarray}\label{r finito}
\frac{b(r)}{r} \leq 1,
\end{eqnarray}
(for $r \geq b_0$) throughout the space--time. This is required in
order to ensure the finiteness of the proper radial distance
$l(r)$ defined by
\begin{eqnarray}
l(r)=\pm \int^r_{b_0} \frac{dr}{\sqrt{1-b(r)/r}}.
\end{eqnarray}
The $\pm$ signs refer to the two asymptotically flat regions which
are connected by the wormhole. The equality sign in~(\ref{r
finito}) holds only at the throat.

Constraint 4: Asymptotic flatness condition, i.e. as $l
\rightarrow \pm \infty$ (or equivalently, $r \rightarrow \infty$)
then $b(r)/r \rightarrow 0$.

Notice that these constraints provide a minimum set of conditions
which lead, through an analysis of the embedding of the spacelike
slice of~(\ref{4wormhole}) in a Euclidean space, to a geometry
featuring two asymptotically flat regions connected by a
bridge~\cite{Dadhich}.

In this paper we are not including considerations about the
traversability constraints discussed by Morris and
Thorne~\cite{Morris}.

\subsection{Self--dual Lorentzian wormholes}
In order to construct a wormhole, one has to specify or determine
the red--shift function $\phi(r)$ and the shape function $b(r)$.
In general one of them is chosen by fiat and the other is
determined by implementing some physical condition. In this paper
we shall use the definition of self--dual wormholes developed
in~\cite{Dadhich} and use it to obtain the n--dimensional
self--dual version. The authors of Ref.~\cite{Dadhich} have
proposed the following equation as the equation for wormhole:
\begin{eqnarray}\label{condicion de dadhich}
\rho=\rho_t=0,
\end{eqnarray}
where $\rho=T_{\alpha \beta}u^{\alpha}u^{\beta}$ and
$\rho_t=(T_{\alpha \beta}-\frac{1}{2}Tg_{\alpha
\beta})u^{\alpha}u^{\beta}$ are the energy density measured by a
static observer and the convergence density felt by a timelike
congruence respectively ($u^{\alpha}u_{\alpha}=-1$).

From Eq.~(\ref{condicion de dadhich}) we have that $T=0$. This
specific restriction on the form of the energy--momentum tensor
automatically leads to a class of wormhole solutions. Since we are
interested in Einstein wormholes, we conclude that the scalar
curvature $R=0$. This constraint will be a condition on the shape
function $b(r)$ and on the red--shift function $\phi(r)$ and on
their derivatives. For obtaining the n--dimensional self--dual
wormhole solution, we first demand $\rho=0$ from which we obtain
$b(r)$, and then solve $T=R=0$ which would determine the function
$\phi(r)$. It is interesting to note that the most general
solution of Eq.~(\ref{condicion de dadhich}) automatically
incorporates the requirement of the existence of a throat without
horizons, i.e. Constraint 1 enumerated above is satisfied.

\section{n--dimensional self--dual Lorentzian wormholes}

\subsection{General $R=0$ solution}

For the spherically symmetric wormhole in higher dimensions we
shall consider the spacetime given by
\begin{eqnarray}
\label{ndim wormhole}
ds^2=-e^{2\phi(r)}dt^2+\frac{dr^2}{1-\frac{b(r)}{r}}+r^2
d\Omega_{n-2}^2,
\end{eqnarray}
where now
\begin{eqnarray}
d \Omega_{n-2}^2=d \theta_2^2+sin^2 \theta_2 d \theta_3^2+.....+
\prod \limits^{n-2}_{i=3} sin^2 \theta_i d \theta^2_{n-1}.
\end{eqnarray}
In other words we have replaced the two--sphere of
metric~(\ref{4wormhole}) by a (n-2)--sphere. In our notation the
angular coordinates are denoted by $\theta_i$, where $i$ runs from
2 to $(n-1)$ and $n$ is the dimensionality of the space-time. Note
that in this notation there is no angular coordinate $\theta_1$.

We now introduce the proper orthonormal basis as
\begin{eqnarray}
ds^2= - \theta^{(0)}
\theta^{(0)}+\theta^{(1)}\theta^{(1)}+\theta^{(2)}\theta^{(2)}+
\nonumber \\ \theta^{(3)}\theta^{(3)} + \sum_{i=4}^{n-1} \,
\theta^{(i)}\theta^{(i)},
\end{eqnarray}
where the basis one--forms $\theta^{(\alpha)}$ (here the index
$\alpha$ runs from $0$ to $(n-1)$) are given by
\begin{eqnarray*}
\theta^{(0)}=e^{\phi(r)} \, dt,
\end{eqnarray*}
\begin{eqnarray*}
\theta^{(1)}= \frac{dr}{\sqrt{1-\frac{b(r)}{r}}},
\end{eqnarray*}
\begin{eqnarray*}
\theta^{(2)}=r d \theta_2,
\end{eqnarray*}
\begin{eqnarray*}
\theta^{(3)}= r \, sin \, \theta_2 \,d \theta_3, . . . . .,
\end{eqnarray*}
\begin{eqnarray*}
\theta^{(n-1)}&=& r \, \prod \limits^{n-2}_{i=2} \,sin \, \theta_i
\,d \theta_{n-1}.
\end{eqnarray*}

From these expressions we obtain for the non--null components of
the Einstein tensor:
\begin{eqnarray}
G_{(0)(0)} =-\frac{(n-2)}{2r^3}\left[\frac{}{} (n-4)b+rb^{\prime}
\frac{}{}\right],
\end{eqnarray}
\begin{eqnarray}
G_{(1)(1)} =\frac{(n-2)}{2r^3}\left[\frac{}{} (n-3)b +
2r\phi^{\prime}(b-r) \frac{}{}\right],
\end{eqnarray}
\begin{eqnarray}
G_{(2)(2)} =\frac{1}{2r^3} \left[ \frac{}{} 2r^2(b-r)(\phi^{\prime
\prime}+\phi^{\prime 2}) \right. \nonumber
\\ \left. +r(rb^{\prime}-7b-2nr+2nb+6r)\phi^{\prime}+b(15-8n) \right. \nonumber \\
\left. +r b^{\prime} (n-3)+n^2 b\frac{}{} \right],
\end{eqnarray}
and
\begin{eqnarray}
G_{(3)(3)} = G_{(4)(4)} =.....= G_{(n-1)(n-1)}= G_{(2)(2)},
\end{eqnarray}
where the prime denotes a derivative with respect to r.

The curvature scalar is given by
\begin{eqnarray}
R=\frac{n-2}{2r^3}\left[\frac{}{} 2r^2(b-r)(\phi^{\prime
\prime}+\phi^{\prime 2})+ \left( \frac{}{} r^2b^{\prime}+2r^2(2-n)
\right. \right. \nonumber \\ \left. \left. +rb(2n-5)\frac{}{}
\right) \phi^{\prime}+(n-4)(n-2)b + rb^{\prime}(n-2)
\frac{}{}\right]. \nonumber \\
\end{eqnarray}

We define the diagonal energy--momentum tensor components as
\begin{eqnarray}
T_{(0)(0)}=\rho(r), T_{(1)(1)}=\tau(r), \nonumber \\
T_{(2)(2)}=T_{(3)(3)}=.....=T_{(n-1)(n-1)}=p(r).
\end{eqnarray}
Here $T_{(2)(2)}=T_{(3)(3)}=.....=T_{(n-1)(n-1)}$, in view that we
have a spherically symmetric space--time.

Using the Einstein equations
\begin{eqnarray}
G_{(\alpha)(\beta)}=R_{(\alpha)(\beta)}-\frac{R}{2}
g_{(\alpha)(\beta)}= - \kappa T_{(\alpha)(\beta)},
\end{eqnarray}
where $\kappa=8 \pi G$, we find from~(\ref{ndim wormhole}) that
\begin{eqnarray}
\label{ndimrho} \kappa \rho(r) = \frac{(n-2)}{2r^3}\left[\frac{}{}
(n-4)b+rb^{\prime} \frac{}{}\right],
\end{eqnarray}
\begin{eqnarray}
\label{ndimtau} \kappa \tau(r) =-\frac{(n-2)}{2r^3}\left[\frac{}{}
(n-3)b + 2r\phi^{\prime}(b-r) \frac{}{}\right],
\end{eqnarray}
\begin{eqnarray}\label{ndimp}
\kappa p(r) =-\frac{1}{2r^3} \left[ \frac{}{}
2r^2(b-r)(\phi^{\prime \prime}+\phi^{\prime 2}) \right. \nonumber
\\ \left. +r(rb^{\prime}-7b-2nr+2nb+6r)\phi^{\prime}+b(15-8n) \right. \nonumber \\
\left. +r b^{\prime} (n-3)+n^2 b\frac{}{} \right].
\end{eqnarray}
Here $\tau(r)$ is the radial pressure $p_r$ (and differs by a
minus sign from the conventions of Morris and
Thorne~\cite{Morris}) and $p(r)$ is the transverse pressure $p_t$.

Considering the condition of self--duality $R=T=0$ we find the
following equation:
\begin{eqnarray}
\label{ecuacion de traza nula} \phi^{\prime \prime}+\phi^{\prime
2}+ \left( \frac{r b^{\prime}+b(2n-5)+2r(2-n)}{2r (b-r)}
\right)\phi^{\prime} \nonumber \\
 +\frac{(n-4)(n-2)b+rb^{\prime}(n-2)}{2r^2(b-r)}=0.
\end{eqnarray}
This equation can be thought  of as the master equation for all
n--dimensional static spherically symmetric $R=0$ geometries.

Given the function $b(r)$ one can solve equation~(\ref{ecuacion de
traza nula}) to obtain $\phi(r)$. We shall find the function
$b(r)$ from the self--dual condition. This implies that
$\rho(r)=0$ and then, from Eq.~(\ref{ndimrho}), we have
\begin{eqnarray}
b(r)=2 m r^{4-n}.
\end{eqnarray}
Here we have written the constant of integration as $2 m$ in order
to obtain the mass $m$ in the four dimensional Schwarzschild case.
Then Eq.~(\ref{ecuacion de traza nula}) simplifies to
\begin{eqnarray}
\label{ecuacion de traza nula con b(r)} \phi^{\prime
\prime}+\phi^{\prime 2}+ \left( \frac{n m r^3-m
r^3+2r^n-nr^n}{2mr^4-r^{n+1}} \right)\phi^{\prime}=0.
\end{eqnarray}

The solution of this nonlinear differential equation may be
written as
\begin{eqnarray}\label{solucionphi}
\phi(r)=ln\left(k+\lambda \sqrt{1-\frac{2m}{r^{n-3}}} \right),
\end{eqnarray}
where $k$ and $\lambda$ are constants of integration. This implies
that the line element~(\ref{ndim wormhole}) takes the form
\begin{eqnarray}\label{solucion final}
ds^2=-\left( k+\lambda \sqrt{1-\frac{2m}{r^{n-3}}}\right)^2 dt^2
\nonumber \\ +\frac{dr^2}{1-\frac{2m}{r^{n-3}}}+r^2 d
\Omega^2_{n-2}.
\end{eqnarray}

The components of the energy--momentum tensor are given by
\begin{eqnarray} \label{Wdensidad}
\rho(r)=0,
\end{eqnarray}
\begin{eqnarray}\label{Wtau}
\kappa \tau(r)=-\frac{k (n-3)(n-2)\,m \, r^{1-n}}{k+\lambda
\sqrt{1-\frac{2m}{r^{n-3}}}},
\end{eqnarray}
and
\begin{eqnarray}\label{Wp}
\kappa p(r)= \frac{k (n-3) \, m \, r^{1-n}}{k+\lambda
\sqrt{1-\frac{2m}{r^{n-3}}}}.
\end{eqnarray}
These expressions satisfy the constraint for the trace of the
energy--momentum tensor:
\begin{eqnarray*}
T=T_{(\alpha)(\beta)}g^{(\alpha)(\beta)}=-\rho(r)+\tau(r)+(n-2) \,
p(r)=0.
\end{eqnarray*}
Note that from equations~(\ref{solucion final})--(\ref{Wp}), for
$n=4$ we obtain the 4--dimensional expressions (9), (10), (11) and
(12) of Ref.~\cite{Dadhich}.

For $n=3$ the Eq.~(\ref{solucionphi}) does not represent the
general solution for the differential equation~(\ref{ecuacion de
traza nula con b(r)}), since the solution~(\ref{solucionphi})
depends on only one integration constant. The general solution now
will be given by
\begin{eqnarray}\label{solpart}
ds^2=-(k+\lambda\, \ln r )^2dt^2+\frac{dr^2}{1-2m}+r^2d\theta^2,
\end{eqnarray}
where $k$ and $\lambda$ are the constants of integration and
\begin{eqnarray*}
\rho=0,
\end{eqnarray*}
\begin{eqnarray*}
\kappa \tau(r) =-\kappa p(r) =-\frac{\lambda(1-2m)}{(k+\lambda \,
\ln r)r^2},
\end{eqnarray*}
are the components of the energy--momentum tensor. The resulting
spacetime is singular at $r_1=0$ and $r_2=e^{-k/\lambda}$. If $k
>> 0$ or $\lambda\approx 0$, then the singular radius $r_2 \rightarrow 0$.

If we set $\lambda=0$, we have from~(\ref{solpart}) a vacuum
solution which is locally flat, but conical in form. This metric
is analogous to the Schwarzschild metric around a point mass in
four dimensions. However, this metric does not lead to a black
hole~\cite{Star}, and then also there is no three--dimensional
Schwarzschild wormhole.

For $n=2$ the metric~(\ref{solucion final}) is not the general
self--dual solution, since in (1+1)--gravity the Einstein tensor
vanishes identically, so that the self--dual
condition~(\ref{condicion de dadhich}) is identically satisfied.
%In particular, the line element obtained from the
%metric~(\ref{solucion final}), for $n=2$,} does not satisfy the
%constraints 2, 3 and 4.

Thus, in the following subsection we shall consider that $n \geq
4$.

\subsection{Black holes, naked singularities and wormholes}
Within the framework of the obtained family of solutions there are
three classes of spherically symmetric spacetimes: black hole,
nakedly singular and wormhole geometries. For example, the
n--dimensional vacuum Schwarzschild geometry~\cite{Myers} is
obtained when $k=0$. If $m < 0$, we have a naked singularity (for
any value of $k$ and of $\lambda$). The n--dimensional
spatial--Schwarzschild traversable wormhole is obtained when
$\lambda=0$.

We are interested in the interpretation of the obtained solution
as a Lorentzian wormhole. The weak energy condition ($\rho \geq
0$, $\rho+\tau \geq 0$, $\rho+p \geq 0$) and null energy condition
($\rho+\tau \geq 0$, $\rho+p \geq 0$) are both violated by
Eqs.~(\ref{Wdensidad})--(\ref{Wp}). Effectively, we have here that
$\rho=0$; then we obtain $\tau \geq 0$ and $p \geq 0$. But $T=0$
implies that $\tau=-2p$, thus, if $p \geq 0$, we have $\tau \leq
0$ (and vice-versa). The extent of the energy condition violation,
caused by the behavior of $r^{1-n}$, for $n \geq 4$, is large in
the vicinity of the throat. One does have a control parameter $k$
which can be chosen  to be very small in order to restrict the
amount of violation.

In the following, we shall follow the procedure of Dadhich et
al.~\cite{Dadhich}. To really make the wormhole explicit we need
two coordinate patches:
\begin{eqnarray}\label{solucion final1}
ds^2=-\left(k+\lambda \sqrt{1-\frac{2m}{r_1^{n-3}}}\right)^2 dt^2
\nonumber \\ +\frac{dr_1^2}{1-\frac{2m}{r_1^{n-3}}}+r_1^2 d
\Omega^2_{n-2},
\end{eqnarray}
where $(2m)^{1/(n-3)} \leq r_1 \leq\infty$ and
\begin{eqnarray}\label{solucion final2}
ds^2=-\left( k-\lambda \sqrt{1-\frac{2m}{r_2^{n-3}}}\right)^2 dt^2
\nonumber \\ +\frac{dr_2^2}{1-\frac{2m}{r_2^{n-3}}}+r_2^2 d
\Omega^2_{n-2},
\end{eqnarray}
where $(2m)^{1/(n-3)} \leq r_2 \leq\infty$.

We have to sew these patches together at
$r=r_1=r_2=(2m)^{1/(n-3)}$. To make this clearer, it is convenient
to go to isotropic coordinates. The advantage of these coordinates
is that in almost all cases a single coordinate patch covers the
entire space--time.

In isotropic coordinates, the metric in a slice $t=$ constant is
conformal to the metric of Euclidean (n-1)--space. We consider a
transformation in which the angular coordinates $\theta_2$ ...
$\theta_{n-1}$ and $t$ remain unchanged while the radial
coordinate
\begin{eqnarray}
r \rightarrow \overline{r}=\overline{r}(r),
\end{eqnarray}
so that $\overline{r}$ is some other radial coordinate, where the
metric~(\ref{solucion final}) takes the form
\begin{eqnarray}
ds^2=-\left( k+\lambda \sqrt{1-\frac{2m}{r^{n-3}}}\right)^2 dt^2
\nonumber \\ + \lambda^2(\overline{r})[d\overline{r}^2 +
\overline{r}^2 d \Omega^2_{n-2}].
\end{eqnarray}

Comparing this metric with~(\ref{solucion final}) we obtain
$r^2=\lambda^2(\overline{r}) \, \overline{r}^2$ and the following
ordinary differential equation:
\begin{eqnarray}
\frac{dr}{\sqrt{r^2-2mr^{5-n}}}= \pm
\frac{d\overline{r}}{\overline{r}}.
\end{eqnarray}
Since we require $\overline{r} \rightarrow \infty$ as $r
\rightarrow \infty$, we take the plus sign. Integrating the latter
expression we find
\begin{eqnarray}\label{transformation}
r=\overline{r} \, \left[ 2^{n-4}\left(1+\frac{m}{(2
\overline{r})^{n-3}} \right)^2 \right]^{1/(n-3)}.
\end{eqnarray}
Note that, when $\overline{r}\rightarrow 0$ or
$\overline{r}\rightarrow \infty$, the coordinate $r \rightarrow
\infty$ for $n \geq 4$.

When $n=4$, we obtain exactly the same transformation for going
from curvature coordinates to isotropic coordinates  as was used
for Schwarzschild  itself. In this case of course the
transformation~(\ref{transformation})  carries the n--dimensional
Schwarzschild metric into isotropic form, since the space part of
the metric~(\ref{solucion final}) is identical to the space part
of the higher dimensional Schwarzschild version.

Thus by~(\ref{solucion final}) and~(\ref{transformation}) the line
element, in terms of isotropic coordinates, is given by
\begin{eqnarray}\label{isotropic form}
ds^2=-\left( k+\lambda
\left[\frac{\frac{m}{(2\overline{r})^{n-3}}-1}{\frac{m}{(2\overline{r})^{n-3}}+1}
\right]\right)^2 dt^2 \nonumber
\\ + \left[2^{n-4} \left(1+\frac{m}{(2 \overline{r})^{n-3}}
\right)^2 \right]^{2/(n-3)} [d\overline{r}^2 + \overline{r}^2 d
\Omega^2_{n-2}].
\end{eqnarray}
Note that from this equation for $n=4$, we obtain the
4--dimensional expression (15) of Ref.~\cite{Dadhich}.

It can be shown that both metrics~(\ref{solucion final1})
and~(\ref{solucion final2}) take the isotropic
form~(\ref{isotropic form}) with the help of
transformation~(\ref{transformation}). Then we have a single
global coordinate chart for the alleged traversable Lorentzian
wormhole. In this case $r\approx 0$ is one of the asymptotically
flat regions and $r\approx \infty$ is the other one.

In order to have a traversable wormhole, we need to know the
ranges of the parameters $k$ and $\lambda$ for which we have no
horizon, i.e. the $g_{t t}$ component does not go to zero. We
suppose $m > 0$, in order to have no naked singularities. There is
a possible horizon when $g_{tt}=0$, thus
\begin{eqnarray}
k\left(\frac{m}{(2\overline{r})^{n-3}}+1\right)+\lambda
\left(\frac{m}{(2\overline{r})^{n-3}}-1 \right)=0.
\end{eqnarray}

Solving this equation we obtain
\begin{eqnarray}
\overline{r}_{_{H}}=\frac{1}{2}\left[\frac{m(\lambda-k)}{\lambda+k}
\right]^{1/(n-3)}.
\end{eqnarray}
In order to have a horizon, the radius $\overline{r}_{_{H}}$ must
be positive; then we have
\begin{eqnarray}
\frac{\lambda-k}{\lambda+k} \geq 0.
\end{eqnarray}
One can show that this horizon is actually a naked curvature
singularity.

To see this more clearly, we calculate the radial and transverse
pressure:
\begin{eqnarray}
\kappa \tau=-\kappa (n-2) p= \nonumber \\  - \frac{k (n-3)(n-2)m
\left[(2\overline{r})^{n-3} + m
\right]^{2(1-n)/(n-3)}}{2^{(1-n)/(n-3)}(2\overline{r})^{1-n}
\sqrt{g_{tt}}},
\end{eqnarray}\\
and we conclude that both pressures diverge as $g_{tt} \rightarrow
0$. Thus, if the parameters $\lambda$ and $k$ satisfy the
inequalities
\begin{eqnarray}\label{in1}
\lambda +k >0 \,\,\,\,\,\,\,\,\,\, and \,\,\,\,\,\,\,\,\,\,
\lambda -k >0,
\end{eqnarray}
or
\begin{eqnarray}\label{in2}
\lambda +k <0 \,\,\,\,\,\,\,\,\,\, and \,\,\,\,\,\,\,\,\,\,
\lambda -k <0;
\end{eqnarray}
we have naked singularities. If the parameters $\lambda$ and $k$
do not satisfy Eqs.~(\ref{in1}) nor~(\ref{in2}), then Constraint 1
enumerated above is satisfied, and curvature singularities do not
form, the $g_{tt}$ components of the metric never go to zero, and
we have a traversable wormhole.

We enumerate now some other properties of the obtained geometry:

(1) The geometry is invariant under simultaneous sign flips
$\lambda \rightarrow - \lambda$ and $k \rightarrow - k$,
independently of the dimensions of the space--time.

(2) $k=0$, $\lambda \neq 0$ is the n--dimensional Schwarzschild
geometry; it is a non--traversable wormhole (in this case the
isotropic coordinate does not cover the entire manifold).

(3) $\lambda=0$, $k \neq 0$ is the zero--tidal--force
spatial--Schwarzschild traversable wormhole (the manifold is
covered completely by the isotropic coordinate system)

(4) $\lambda=0$, $k=0$ is singular, independently of the
dimensions of the space--time.

(5) At the throat $g_{tt}(r=[2m]^{1/(n-3)})=-k^2$, so $k \neq 0$
is required to ensure traversability.

(6) $g_{tt}(\overline{r}=\infty)=-(k-\lambda)^2$ and
$g_{tt}(\overline{r}=0)=-(k+\lambda)^2$; then we see that time
runs at different rates in the two asymptotic regions.

\section{Concluding remarks}
Our aim in this paper has been to obtain self--dual Lorentzian
wormholes in the framework of n--dimensional Einstein gravity. For
this we have followed the prescription provided by Dadhich et
al.~\cite{Dadhich} for obtaining wormholes. They have proposed one
such prescription which is characterized by the equation
$\rho=\rho_t=0$, which implies that $R=0$ and, equivalently, a
traceless energy--momentum tensor. This constraint gives a
condition on the shape function $b(r)$ and on the red--shift
function $\phi(r)$ and on their derivatives.

For obtaining the n--dimensional self--dual wormhole solution, we
first demand $\rho=0$ from which we obtain $b(r)$, and then solve
$T=R=0$ which would determine the function $\phi(r)$.

The resulting geometry represents a family of spacetimes which
contains Lorentzian wormholes, naked singularities, and the
n--dimensional Schwarzschild black hole. However, in lower
dimensional Einstein gravity, without cosmological constant, we
can not have self--dual Lorentzian wormholes.

The isotropic coordinates are used to display the full structure
of the obtained higher dimensional geometry.

In this case, of course the wormhole geometry has exotic matter as
the source, i.e. the energy--momentum tensor of the matter
explicitly violates the energy conditions. This fact can not be
used to rule out wormhole solutions. Today many physical
situations are known in which the energy conditions are
violated~\cite{Dadhich,Visser1}.

\section{Acknowledgements}
The authors thank Carol Mu\~noz for typing this manuscript and S.
del Campo, N. Cruz and S. Lepe for valuable comments. This work
was supported by CONICYT through Grant FONDECYT N$^0$ 1010485 and
by Direcci\'on de Promoci\'on y Desarrollo de la Universidad del
B\'\i o--B\'\i o.

\end{document}